\documentclass[lettersize,journal]{IEEEtran}
\usepackage{amsmath,amsfonts}
\usepackage{algorithmic}
\usepackage{algorithm}
\usepackage{array}
\usepackage[caption=false,font=normalsize,labelfont=sf,textfont=sf]{subfig}
\usepackage{textcomp}
\usepackage{stfloats}
\usepackage{url}
\usepackage{verbatim}
\usepackage{graphicx}
\usepackage{cite}
\usepackage{titlesec}
\titlespacing*{\subsection}
{0pt}{0.35\baselineskip}{0.35\baselineskip}
\titlespacing*{\section}
{0pt}{0.35\baselineskip}{0.35\baselineskip}
\usepackage[belowskip=-15pt,aboveskip=5pt,font={footnotesize}]{caption}

\begin{document}

\title{Maintaining reliability while navigating unprecedented uncertainty: a synthesis of and guide to advances in electric sector resource adequacy}

\author{Gabriel Mantegna, Ziting Huang, Guillaume Van Caelenberg, Bethany Frew, Muireann Lynch, Mark O'Malley
\thanks{Mr. Mantegna is with the Andlinger Center for Energy and the Environment, Princeton University, Princeton, NJ, USA.}
\thanks{Ms. Huang is with the Department of Environmental Health and Engineering, Johns Hopkins University, Baltimore, MD, USA}
\thanks{Mr. Van Caelenberg and Dr. O'Malley are with the Department of Electrical and Electronic Engineering, Imperial College London, London, UK.}
\thanks{Dr. Frew is with the National Renewable Energy Laboratory, Golden, CO, USA.}
\thanks{Dr. Lynch is with the Economic and Social Research Institute, Dublin, IE.}
\vspace{-1cm}}

\markboth{Pre-print submitted to IEEE Transactions on Energy Markets, Policy and Regulation}{}

\maketitle

\begin{abstract}
The reliability of the electric grid has in recent years become a larger concern for regulators, planners, and consumers due to several high-impact outage events, as well as the potential for even more impactful events in the future. These concerns are largely the result of decades-old resource adequacy (RA) planning frameworks being insufficiently adapted to the current types of uncertainty faced by planners, including many sources of deep uncertainty for which probability distributions cannot be defensibly assigned. There are emerging methodologies for dealing with these new types of uncertainty in RA assessment and procurement frameworks, but their adoption has been hindered by the lack of consistent understanding of terminology related to RA and the related concept of resilience, as well as a lack of syntheses of such available methodologies. Here we provide an overview of RA and its relationship to resilience, a summary of available methods for dealing with emerging types of uncertainty faced by RA assessment, and an an overview of procurement methodologies for operationalizing RA in the context of these types of uncertainty. This paper provides a synthesis and guide for both researchers and practitioners seeking to navigate a new, much more uncertain era of power system planning.
\end{abstract}


\section{Introduction}
\begin{figure}[t]
  \centering
  \includegraphics[width=3.5in]{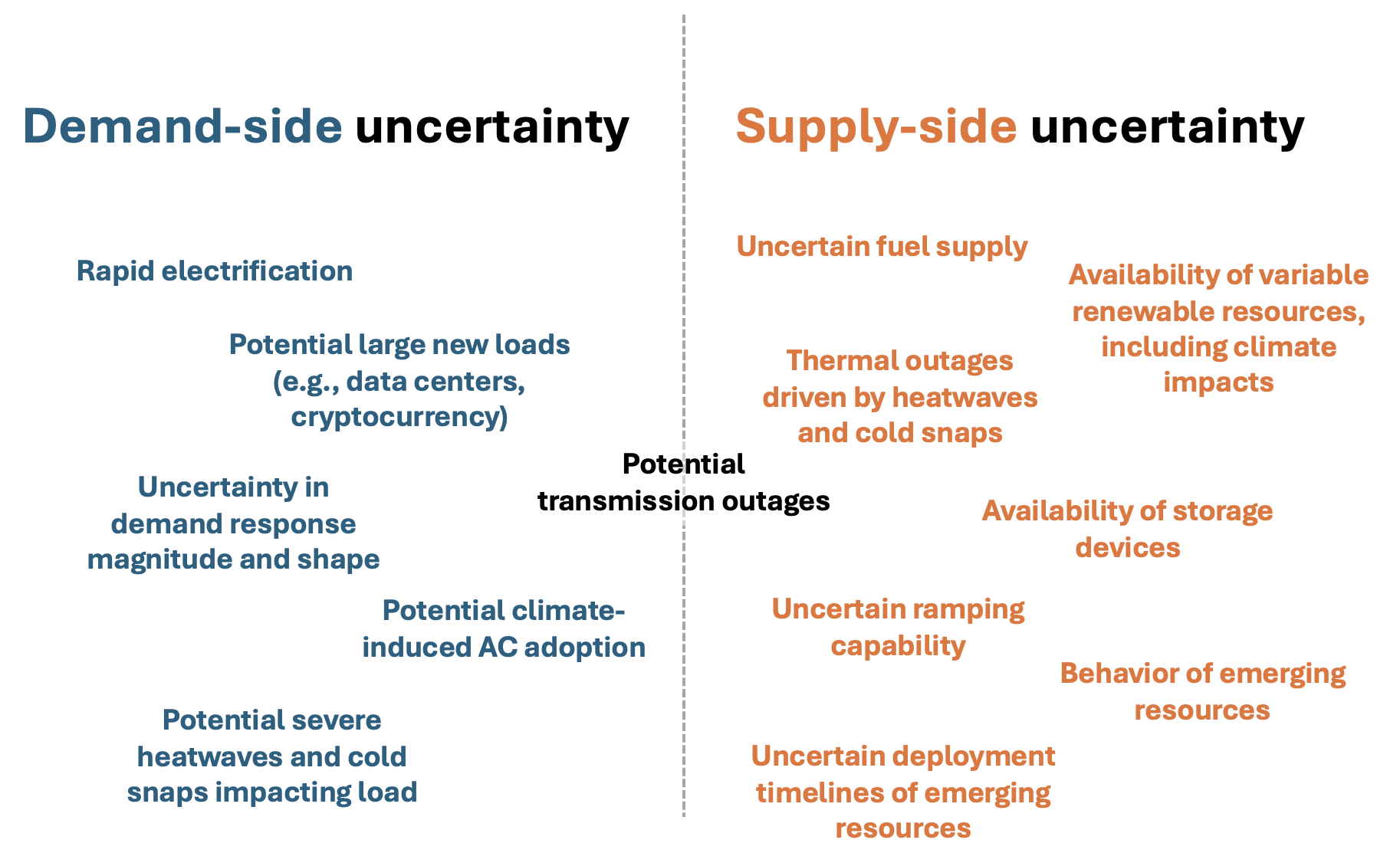}
  \caption{A schematic of the various types of uncertainty affecting power system reliability, organized into uncertainty affecting both the supply and demand sides.}
  \label{fig:unprecedented_uncertainty}
\end{figure}

\IEEEPARstart{T}{he} reliability of the electric grid, while overall extremely high in the developed world, has in recent years become a larger concern for regulators, planners, and consumers due to several high impact events and the potential for even more impactful events in the future. Some salient examples are the multi-day severe outage event that occurred in Texas in February 2021 \cite{ferc_february_2021}, the rotating outages that occurred in California in August 2020 \cite{california_iso_root_2021}, the winter energy crisis in Europe following the invasion of Ukraine  \cite{cuff_europe_2023}, and the large-scale power shortage events and provincial-level rolling blackouts in China between 2020 and 2022 \cite{shen_coping_2024}. These were diverse events with different causes, but what they had in common is an attribution to reliability assessment and planning frameworks that were not designed to handle the broad range of types of uncertainty that we face in the context of modern power system planning. Figure \ref{fig:unprecedented_uncertainty} shows that these uncertainties can generally be classified into uncertainty surrounding the supply and demand sides, and the transmission network that links them, and range from matters as diverse as rapid electrification to geopolitical changes affecting fuel supply.

In tandem with this increasing range of uncertainties that modern planning frameworks must contend with, the reliability of the electric grid is expected to become more important to society than ever. As more end uses are electrified in support of climate change mitigation, modern societies are becoming more dependent on the electric grid, making the consequences of outages potentially worse. Therefore, as we become more dependent on electricity sourced from variable renewables, and as climate change-driven extremes increase, uncertainty will become more important threatening the reliability of the electric grid.

These emerging concerns are largely the result of planning and procurement frameworks that have not evolved to meet the needs of modern electric grids \cite{electric_power_research_institute_resource_2022-1}. In the mid-20th century when the developed world was just starting to design electric grids that could supply demand at nearly all times, the chief reliability concern was generator and transmission outages happening during times of peak demand, so reliability planning consisted of simply quantifying the probability distributions of both generator forced outages and peak electricity demand, and then calculating how much installed capacity was necessary to ensure a pre-determined level of reliability. Reliability planning has come a long way since these frameworks were first developed, but the reliability planning methodologies in use today still carry significant vestiges of these original practices, that may not be fit for purpose in current and future circumstances. It is not a stretch to summarize recent changes to reliability planning as consisting of small changes at the margin to a capacity-based framework which was originally developed for a very specific set of reliability challenges. The electric grid in most of the developed world is extremely reliable (see, e.g., \cite{national_renewable_energy_laboratory_explained_2024}), so it is clear that the methodologies used are not completely inadequate, but the cracks in these methodologies are starting to show as evidenced by recent events, and it is clear that we need to re-evaluate how we do reliability planning as we look forward to a much more uncertain future. Additionally, it is clear that procurement methodologies, including but not limited to markets, need to evolve in line with these changes.

The purpose of this paper is to address this need for a modernizing of reliability assessment and procurement practices, by providing a synthesis of modern advances in the field in and outside the academic literature, as well as a comprehensive mapping to enable the navigation of the key issues related to resource adequacy, that is aimed to support decision makers such as system operators. This paper was supported and inspired by a series of workshops organized by the Global Power System Transformation Consortium (G-PST) in August and September 2023, in which a range of stakeholders including system operators, academics, and consultants came together to identify state of the art practices and key issues in the field of electric grid resource adequacy.  In this paper, we focus on a subset of electric grid reliability known as \textbf{resource adequacy (RA)}, which we will define in more detail below, but which in short refers to the ability of an electric grid to meet demand at most or all times.

This paper is organized as follows. Section \ref{defn} provides a definition of RA that will be used in the paper, and attempts to clarify common confusion on how RA relates to resilience. Section \ref{uncertainty} provides a taxonomy of different types of uncertainty faced in RA planning, and introduces state of the art practices for handling uncertainty in RA assessments, including quantifying risks resulting from the uncertainty. Section \ref{flexibility} discusses the issue of capturing operational details such as transmission constraints, energy storage operations, and flexible demand, which arises from the presence of structural uncertainty, one of several types of uncertainty introduced. Section \ref{procurement} briefly makes the link of to  procurement of resources in support of RA, and touches on topics relevant to both market and non-market areas. Finally Section \ref{synthesis} synthesizes the findings of the paper, provides a framework for operationalizing the findings, and discusses important future research directions.

\section{Defining RA: what is it and how does it relate to resilience?}
\label{defn}

We begin by providing a working definition of RA that will be used in this paper, along with a brief discussion of how RA relates to resilience. Our motivation is that there have been many recent debates, both in the academic literature and in the grey literature and professional organizations, on what types of conditions and events RA assessments should evaluate, and therefore a significant amount of confusion on what RA refers to. In general, our philosophy is to use definitions that are consistent with those used by reliability organizations (particularly NERC), engineering professional organizations, and the risk analysis literature. With this in mind, we first present our definition of RA, which is: \textbf{resource adequacy, in the context of a power system, is the state of having sufficient supply and/or flexible demand such that involuntary load shed events have a probability of occurrence that is deemed acceptable to the power system's regulator.} In the following sections, we unpack this definition by clarifying what it does and does not include.

\subsection{Distinguishing adequacy from operating reliability}
\label{distinguishing}

We will begin by clarifying what is \textbf{not} included in this definition of RA, which is considerations of operating reliability. To inform this distinction, we quote the North American Electric Reliability Corporation (NERC) definition of reliability, which defines reliability as consisting of two distinct aspects, adequacy (it is from this definition that we develop ours; ours is modified to emphasize the probabilistic nature of RA) and operating reliability \cite{north_american_electric_reliability_corporation_reliability_2013}:

\begin{enumerate}
    \item ``\textbf{Adequacy}: The ability of the electricity system to supply the aggregate electrical demand and energy requirements of the end-use customers at all times, taking into account scheduled and reasonably expected unscheduled outages of system elements.
    \item \textbf{Operating Reliability}: The ability of the Bulk-Power System to withstand sudden disturbances, such as electric short circuits or the unanticipated loss of system elements from credible contingencies, while avoiding uncontrolled cascading blackouts or damage to equipment.''\footnote{Note that operating reliability is also often known as system security.}
\end{enumerate}

These NERC definitions emphasize that adequacy is a fundamentally different concept from operating reliability. The best illustration of this difference is the fact that these two types of reliability have very different failure modes associated with them. When a system is \textit{not adequate}, the failure mode is that load shedding occurs; i.e., the system operator intentionally shuts off certain demand in order to keep supply and demand in balance. The most salient result of a system experiencing a \textit{lack of operating reliability}, on the other hand, is a cascading failure, which most commonly occurs when the electric grid AC frequency drops below the range that power system equipment is designed to operate in, causing a cascade of failures across components. In addition to having different failure modes, these two types of reliability are also characterized by much different timescales and dynamics, as well as much different models that are needed to characterize them.

\subsection{What RA does encompass}

We have now established what RA does not encompass, which is considerations of operating reliability, but it is equally important to emphasize what RA does encompass. RA is a state of a system that means it has sufficient available supply and/or flexible demand to keep the probability of load shed events below a certain threshold. This means that any involuntary load shed event, regardless of cause, is a RA issue and the potential for all such events needs to be evaluated as part of a RA assessment. In particular, there are three main categories of events that can all lead to load shed, and need to be considered in a RA assessment:

\begin{itemize}
    \item \textbf{Insufficient installed capacity}. This can happen when demand is higher than expected, and perhaps also one or more generators (or transmission lines) experience a forced outage. These events are the main failure mode leading to load shedding events in thermal-dominant systems with minimal fuel supply concerns, and are thus the focus of ``traditional'' RA planning \cite{electric_power_research_institute_resource_2022}.
    \item \textbf{Insufficient energy supply to meet demand.} This can happen either when enough energy-limited resources such as variable renewables, hydro, or energy storage have an insufficient amount of energy available to dispatch, or when gas generators experience an interruption in their supply of natural gas. This failure mode is becoming more common due to the increasing dependence on variable renewable energy sources and gas generators \cite{electric_power_research_institute_resource_2022}.
    \item \textbf{Insufficient flexibility.} This can happen either when there is insufficient ramping capacity to meet fast ramps in net load, or when there are insufficient operating reserves available (either committed, or existing on the system) to meet changes in supply or demand. This failure mode is less common, but it is becoming more important as power systems around the world transition to higher levels of variable renewable energy, and it was identified as an important cause of the August 2020 rolling outages in CAISO \cite{california_iso_root_2021}.
\end{itemize}

As these events can all lead to load shedding, they should all be evaluated in RA assessments. The occurrence or nonoccurrence of these events is sometimes referred to using terms like ``capacity adequacy'', ``energy adequacy'' or ``flexibility adequacy'' \cite{celebi_briefing_2024, electric_power_research_institute_resource_2022-1}, but our view is that RA should be the common theme, and modeling framework, that unites all these considerations. As illustrated above by the different kinds of events that can all lead to load shedding, a resource adequate system features \textbf{all} of energy adequacy, capacity adequacy, and flexibility adequacy.

\subsection{The role of probability in RA, and the relation of RA to resilience}

An important feature of our definition of RA is the role of probability-- a system is considered resource adequate if the probability of load shed events is below a certain threshold. This means that we have to know the joint probability distribution of all of a system's underlying components if we are to evaluate the RA of that system. But what if we as planners are worried about events for which we do not know their probability? These events for which we know or roughly know the consequences, but do not know the probability of occurrence, are also known as grey swans, and their presence is an important dilemma facing power system planners around the world. For example: what if a hurricane were to hit New York City? Or, what if Russia were to invade Ukraine? Both of these were events that of course occurred and that significantly impacted energy  supply, and that were foreseen as plausible before their occurrence \cite{lin_physically_2012}, albeit without known probabilities.

\begin{figure}[t]
  \centering
  \includegraphics[width=3.5in]{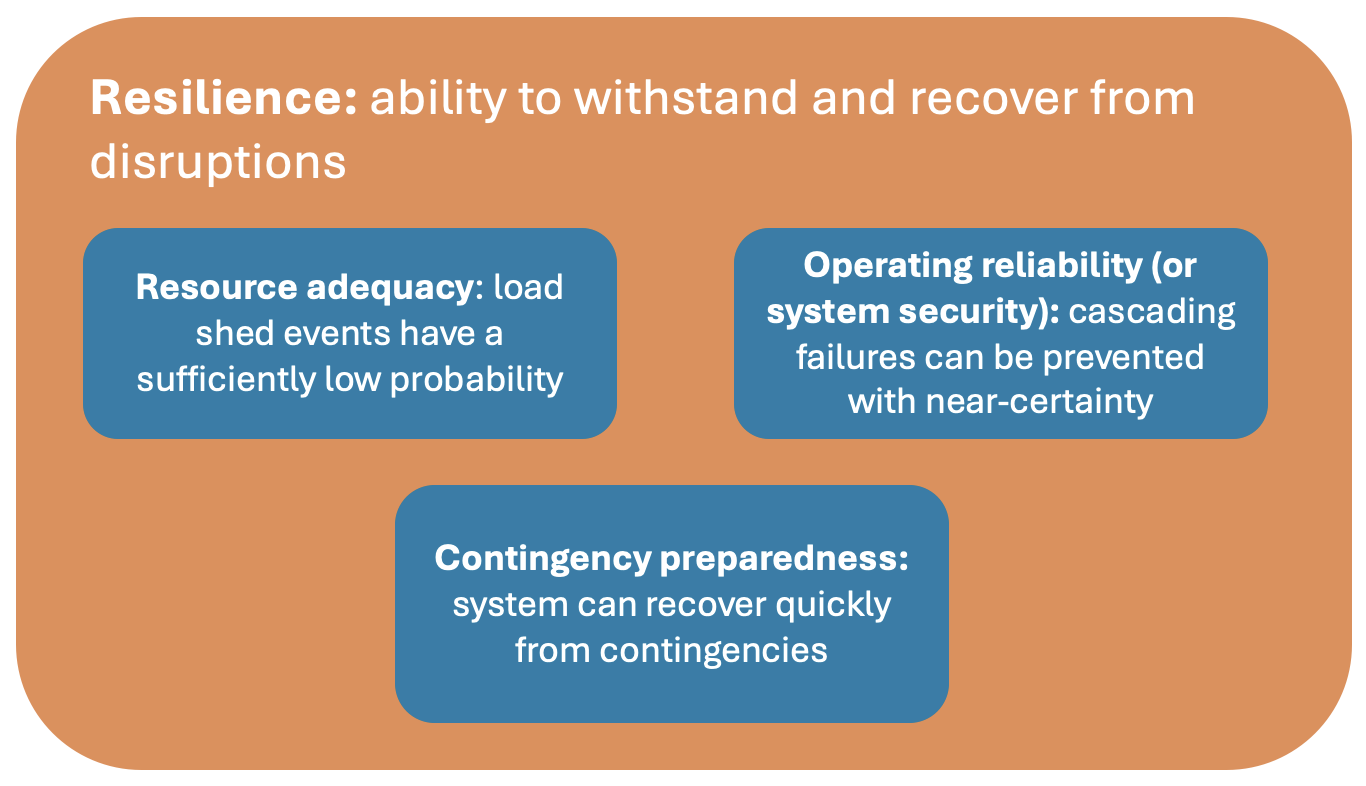}
  \caption{The relationship between resilience, and its key components in the context of the electric grid: RA, operating reliability, and contingency preparedness.}
  \label{fig:resilience}
\end{figure}

This desire for a concept of reliability that is broader and more inclusive of events like grey swans has led to a significant surge of interest across infrastructure planning sub-fields in the concept of resilience \cite{ayyub_climate-resilient_2018, hickford_review_2017, stout_power_2019, fang_adaptive_2019, belle_resilience-based_2023}. This newfound focus on resilience has opened many productive discussions but has also caused some confusion on the relationship between resilience and RA in the context of the power sector. There are many, sometimes competing definitions of resilience, but one convenient and domain-specific one is that used by a recent NREL and USAID report \cite{stout_power_2019}: ``[Power sector resilience is the] ability to anticipate, prepare for, and adapt to changing conditions and withstand, respond to, and recover rapidly from disruptions to the power sector through adaptable and holistic planning and technical solutions.'' There are two general pieces here: the ability to \textbf{withstand} disruptive events, and the ability to \textbf{recover} from disruptions quickly. The general vagueness in this definition has led to disagreement in the literature on whether resilience is a separate concept from reliability and RA, or whether it is simply a broader concept that includes RA \cite{national_academies_of_sciences_engineering_and_medicine_enhancing_2017, kasina_resilience_2024, cainey_resilience_2019}. However, our view, in light of the very broad definition of resilience as being about withstanding changing conditions, is that resilience encompasses the probabilistic notion of reliability, and includes two additional components: 1) preparedness for events for which we do not know the probability, and 2) the mitigation of impacts and ensuring of a quick response if outages do happen. In our view, resilience also encompasses operational reliability in addition to RA, as resilience is about the ability to withstand many kinds of disruptions. The relationship between these terms is summarized in Figure \ref{fig:resilience}.

 In this paper, we focus on laying out best practices for RA assessment and procurement in the probabilistic sense, as opposed to best practices for the broader goal of assessing and planning for resilience. In other words, \textbf{we assume that the goal of RA assessment and planning is to ensure that a system meets a target level of reliability, based on the best available quantifications of the uncertainties involved}. If a planner acknowledges that there are additional uncertainties involved affecting reliability that cannot be quantified (i.e. there is no sound scientific argument and/or data available to quantify them), it is then an implicit assumption that the risk metrics quantified by an RA assessment may be biased, but this can still mean that an RA assessment represents the best available estimation of the risk. The development of best practices for integrated reliability and resilience planning is an area of active research; several recent studies have identified the potential for resilience planning via identification of specific events, and modification of resource plans in order to cost effectively mitigate the events' impacts \cite{kasina_resilience_2024}. At a minimum, if planners are worried about the impact of events for which they cannot confidently assign a probability, a simple first step is to do ``stress tests'' to see the impact of if these events were to happen \cite{electric_power_research_institute_resource_2022-1}. This can be a useful component of RA assessments, but also presents an issue when it comes to planning and procurement because it is likely impossible to evaluate the cost effectiveness of investments to protect against events which have an unknown probability. Relatedly, when it comes to sources of uncertainty which have unknown probabilities, it can also be useful to identify the direction and possibly even rough magnitude of the impact of these sources of uncertainty, to give decision makers an idea of how probabilistic metric may be biased.

 This key role of uncertainty in RA assessment and subsequently procurement motivates the next section, which discusses the different kinds of uncertainty and how they can and should fit into an RA assessment.

 \section{On uncertainty in RA: a taxonomy, and risk metrics}
\label{uncertainty}

RA assessment is fundamentally a problem of quantifying the risk of lost load events, a risk which arises due to the presence of uncertainty in generator availability, network availability, and demand. There are many different types of uncertainty affecting the risk of lost load events, so it is important to understand which types of uncertainty can and should be included in an RA assessment. Therefore, in this section we provide a brief overview of the different types of uncertainty, and use it to inform a discussion of what types of uncertainty can and should be included in RA assessments.

\subsection{A taxonomy of uncertainty}

There are two, independent ``dimensions'' of uncertainty we wish to highlight: 1) the dimension of parametric vs structural uncertainty, and 2) the ``level'' of uncertainty, which represents the extent to which we can quantify the uncertainty. Earlier in Figure \ref{fig:unprecedented_uncertainty} we presented a classification of sources of uncertainty into supply- and demand-side related uncertainties; here we will present a classification of uncertainties that is more consistent with the risk analysis literature and that allows for a clearer connection to how different sources of uncertainty can and should be modeled.

The first dimension is the dimension of distinguishing parametric from structural uncertainty. Parametric uncertainty refers to uncertainty in the parameters of a model; for example, the availability of a generator or the load shape for a particular year. Almost all models and decision problems have parametric uncertainty. Structural uncertainty, on the other hand, refers to uncertainty arising from the structure of a model; i.e. from it being a less-than-perfect representation of the real world. For example, the leaving-out of transmission constraints in an RA model represents structural uncertainty. Structural uncertainty is difficult to quantify, but it is important to understand when it is present as it can be a significant source of uncertainty impacting the conclusions of a modeling exercise. When we do RA assessments, the types of uncertainties we seek to input into our models are always parametric uncertainties, and we generally also seek to minimize structural uncertainty by understanding which components of a model are important to include.

\begin{figure}[t]
  \centering
  \includegraphics[width=3.5in]{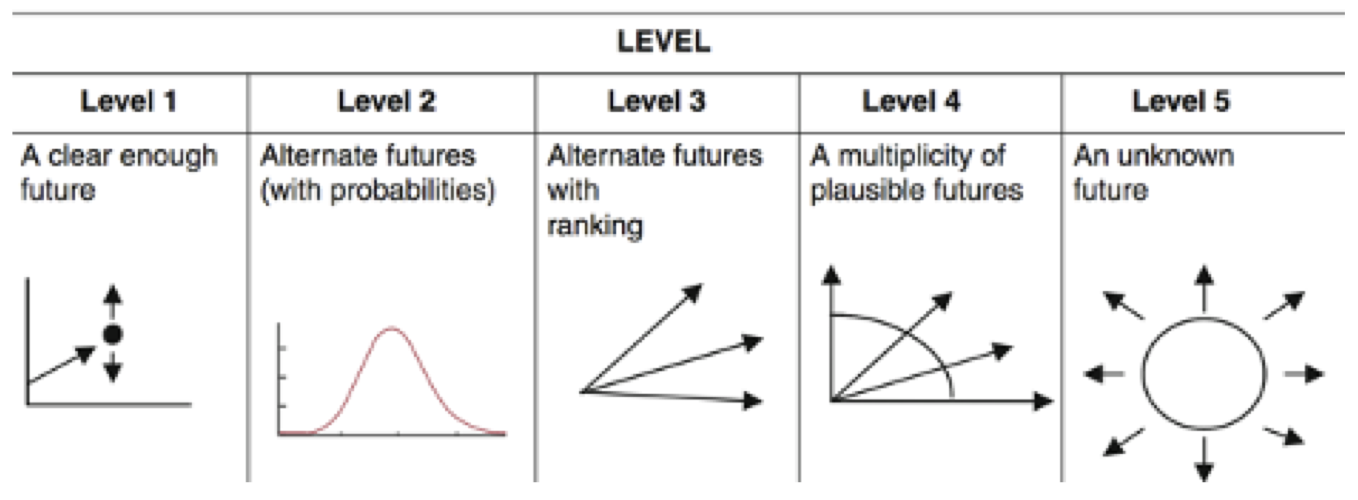}
  \caption{A taxonomy of the 5 levels of parametric uncertainty as conceived by Cox \cite{cox_jr_confronting_2012} and adapted by Thissen \cite{thissen_public_2013}.}
  \label{fig:uncertainty}
\end{figure}

The second dimension is the ``level'' of uncertainty. This dimension applies only to parametric uncertainty and not structural. Per the definition originally proposed by \cite{cox_jr_confronting_2012} and adapted by \cite{thissen_public_2013}, parametric uncertainty can be thought about as having 5 levels, in order from most to least quantifiable. These levels are visualized in Figure \ref{fig:uncertainty}.

\begin{figure}[t]
  \centering
  \includegraphics[width=3.5in]{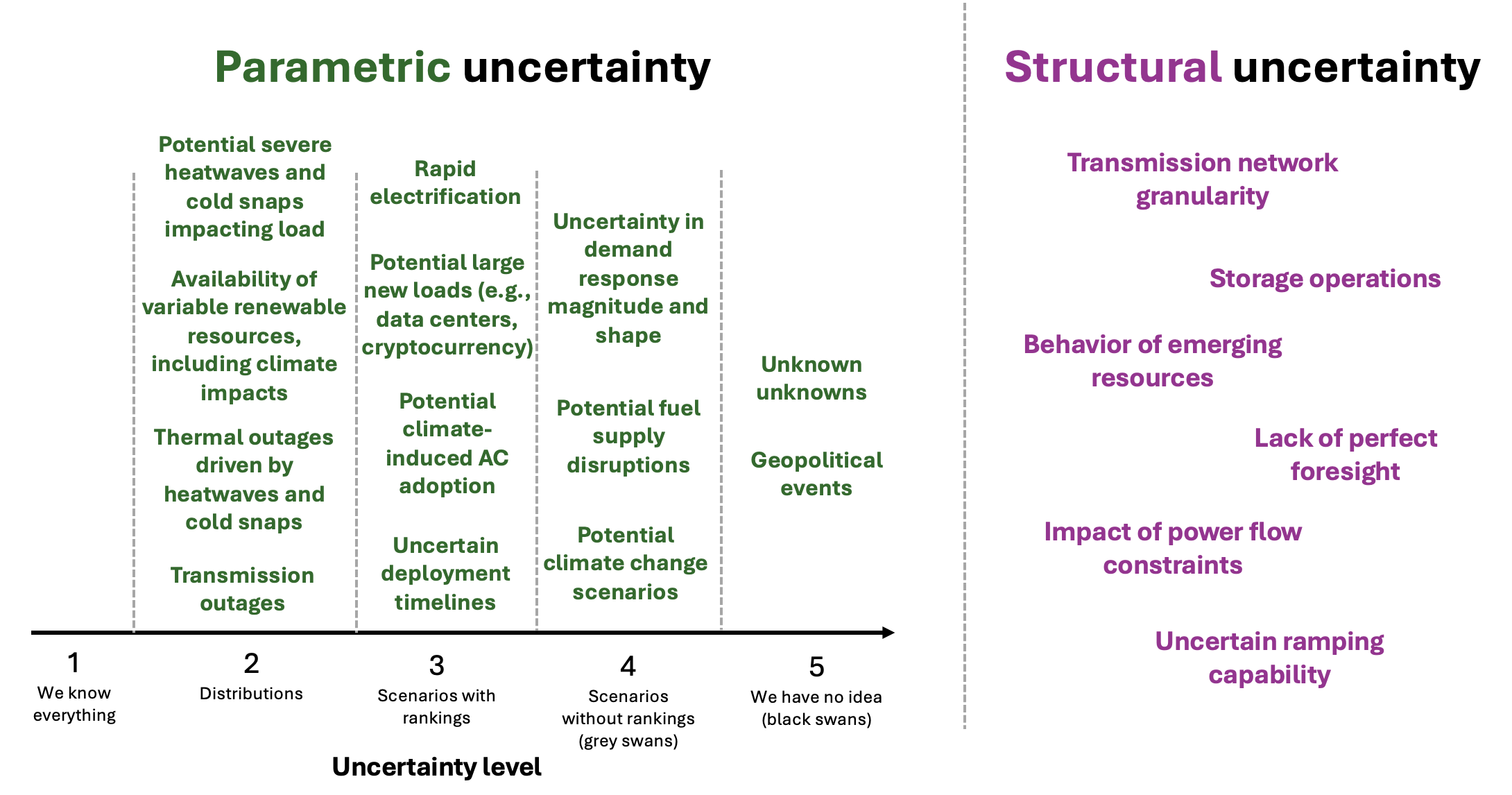}
  \caption{A schematic of the dimensions of uncertainty presented here, along with a depiction of where the key types of uncertainty affecting power system RA fall in this categorization.}
  \label{fig:uncertainty_plot}
\end{figure}

Level 1 uncertainty refers to essentially minimal uncertainty: we have a good estimate of how the system element in question will behave in the future. Level 2 uncertainty refers to uncertainty for which we can quantify a probability distribution. Per our definition of RA as referring to a probabilistic assessment of the risk of rolling outages, all uncertainties that should be inputted into an RA assessment are Level 2. Level 3 uncertainty refers to when we have an idea of the range of possibilities for the system parameter in question, with a ranking of which is most likely. Level 4 uncertainty refers to when we have only an idea of the range of possibilities for a system parameter, without an idea of which is most likely. Events with Level 4 uncertainty surrounding their occurrence are also known as ``grey swans;'' these are events which we know are possible but do not have a way to confidently calculate the probability of. Finally Level 5 uncertainty refers to when we know that a system parameter is uncertain, but do not have an idea of the range of possible outcomes. Events with Level 5 uncertainty surrounding their occurrence are also known as ``black swans;'' these are events which planners do not even foresee as possible beforehand. Levels 3 through 5 are also known as ``deep uncertainty'', which simplify refers to the fact that it is not possible or practical to develop probability distributions associated with them. Figure \ref{fig:uncertainty_plot} depicts where the key sources of uncertainty affecting RA fall along the categorization presented here.

\subsection{Uncertainty in RA assessments}

Ideally, all sources of parametric uncertainty affecting RA could be included in an RA assessment, which would allow for the confident calculation of a ``true'' loss of load probability. In reality, however, many sources of uncertainty are difficult to quantify, and other sources of uncertainty are ``deep'' (levels 3-5), meaning frequentist probabilities simply do not exist to quantify them. Our opinion is that the solution to this problem is to incorporate into RA assessments the best available estimates of probability distributions for all the sources of uncertainty affecting reliability, acknowledging that this may lead to risk metrics that are not ``perfect,'' but which are the best available. As mentioned above, there are emerging methodologies for planning for grey swan events, but they are not yet well developed in the academic literature or in industry (for an example, see \cite{kasina_resilience_2024}). Even if these methodologies do become more developed, however, it will fundamentally never be possible to develop probabilistic reliability metrics which incorporate events for which we do not know the probabilities.  It is our view that probabilistic RA analysis will remain relevant as a key part of reliability and resilience planning but its limitations need to be recognised and brought to bear on any interpretation and/or use e.g. in procurement (see below). In short, RA analysis should consist of calculating the best estimate of loss-of-load risk, and can and should eventually be paired with other frameworks that incorporate planning for grey swans.

With this strategy in mind, we will briefly highlight the components of what we see as a ``best practice'' RA assessment. In an ideal RA assessment for a given portfolio, chronological operations of a power system are simulated for many weather years (ideally but not necessarily using economic dispatch), and a Monte Carlo technique is used to introduce probabilistic variations in load and generator availability, based on the correlations of each with weather conditions. Simulating chronological operations is necessary in order to determine the availability of storage resources; this and other considerations are discussed further below in Section \ref{flexibility}. Thus there are essentially two steps here: 1) developing a time series of weather data (including renewables availability) which represents the range of weather conditions the analyst thinks is plausible for the system, and 2) developing probabilistic representations for outages of system components, importantly including their correlation with weather and with each other. It is important to note that, given the presence of climate change which significantly affects weather conditions, it is preferable that the weather conditions used in an RA simulation are developed using information from climate models, so that they represent the range of weather conditions possible in the future year reliability is being calculated for.

\subsection{Quantifying risk in RA assessments}

Another important question in RA modeling which has received much attention in recent literature is the question of how exactly to evaluate the risk of lost load events in an RA assessment. Due to its thorough treatment in other recent literature (referenced below), we will give only a brief summary. The traditional metric used in RA assessment is Loss-of-Load Probability, or LOLP, which refers to the probability of having a lost load event in a day, no matter the duration or intensity. However, recent literature has established that, particular in the context of renewable-dominated energy systems, it is important to look at more metrics. Loss of load events have not just a frequency of occurrence, but also an \textbf{intensity} and a \textbf{duration} when they occur. Therefore, recent literature has identified Loss-of-Load Expectation (LOLE, referring to duration), and Expected Unserved Energy (EUE, referring to intensity) as other important metrics to examine in RA assessments \cite{electric_power_research_institute_resource_2022, stenclik_redefining_2021}. Some recent reports have also advocated for looking at the distribution of these metrics rather than just their expected value \cite{electric_power_research_institute_resource_2022}. These considerations are important because they allow for planners and decision makers to have a fuller consideration of RA risk and also target mitigation and procurement measures appropriately (Section \ref{procurement} below elaborates more on the question of best practices for procurement).

Another important, related topic is the question of ``how much'' reliability a planner should plan for; i.e. what the target risk should be, which may or may not be the traditional ``1 day in 10 years'' metric. This is separate from the topic of evaluating risk for a particular portfolio of resources, but is an important topic that also involves equity considerations \cite{stenclik_redefining_2021, lin_equity-based_2022}.

\subsection{Modeling extreme events}

There is a tendency across the electric utilities industry to treat extreme weather events as ``force majeure'' events that it is not possible or worthwhile to plan for, and that are somehow separate from probabilistic planning frameworks. This tendency exists in both the distribution grid planning space, where often extreme weather events are excluded from distribution outage metrics, and in the generation and transmission planning space, where extreme weather events are sometimes spoken about as if they are separate from RA planning. We wish to emphasize, as this is a common point of confusion, that while there are certainly sources of uncertainty (e.g., geopolitical changes; see Figure \ref{fig:uncertainty_plot}) for which probabilities cannot generally be defensibly assigned, extreme weather events are best considered within probabilistic RA assessment frameworks to the extent that we have data to represent their frequency of occurrence. Extreme weather events can be considered within RA assessment models through the inclusion of many weather years of weather data in line with the best practices described above. Further, climate change-driven weather extremes can be captured through the use of climate model outputs in lieu of historical weather conditions, although this is not yet common practice. Explicitly modeling weather extremes within RA frameworks is crucial because this allows for risk-informed planning: if a given type of event (e.g., climate change-drive extreme heat waves) is included in an RA assessment probabilistically, then planning frameworks can adapt accordingly based on the frequency of occurrence of these events in order to manage risk, but if this event is only included as a stress test, then it becomes more difficult to manage risk based on objective characterizations. Therefore, events should be ``brought within the RA fold'' when possible, to the extent that we have probabilities to describe them and that they are impactful.

\section{Operational details in RA}
\label{flexibility}

At this point, we have discussed what RA means, discussed the types of uncertainty that can and should be included in RA, and established that RA assessment should consist of the chronological, Monte Carlo-assisted simulation of many possible system conditions for a given portfolio. These sections establish the basic framework of an RA assessment, but there still remains the question of what system details to include in the simulation. Since many (hundreds or thousands) years of system conditions must be simulated to get an accurate picture of risk \cite{pfeifenberger_resource_2013}, RA assessment will be computationally challenged if a detailed production simulation including power flow is employed, so modelers must decide what system details to include, a problem of structural uncertainty. Additionally, there are certain system elements such as energy storage and flexible demand which are both energy limited (in that they cannot dispatch indefinitely), and contain uncertainty regarding their operations, which presents additional structural uncertainty.

These topics are covered extensively in recent literature \cite{leibowicz_importance_2024, electric_power_research_institute_modeling_2023, sun_insights_2022, nolan_capacity_2017}, so in this section we summarize best practices for modeling a) operational constraints, b) energy storage, and c) flexible demand, in a relatively concise manner, encouraging the reader to reference the recent literature on the topic for more details.

\subsection{Operational constraints}

Traditionally, RA assessments take a highly simplified approach to evaluating the risk of lost load, often excluding important operational constraints such as transmission constraints and unit commitment. However it has recently been found by several studies that excluding these factors can bias the risk of lost load substantially (usually, the exclusion of these factors will lead to reliability being overestimated) \cite{leibowicz_importance_2024,sun_insights_2022}. Therefore in this section we draw out best practices regarding the inclusion of these factors.

One important best practice regarding operational constraints in RA assessment is to include an explicit representation of the transmission network, so that the potential for transmission limits and/or outages to contribute to RA issues can be identified \cite{leibowicz_importance_2024}. Assuming no limits or outages will bias the results and over estimate RA.  It is also important to model neighboring regions, so that the potential for imports at crucial hours can be identified. Many RA assessment models used by industry parametrize transmission considerations by de-rating resource availability according to deliverability during historical conditions; however recent work has identified that explicitly modeling the transmission network within an RA model is important for accurately capturing load shed risk \cite{leibowicz_importance_2024} as de-rating is somewhat arbitrary leading to bias.

A second important best practice regarding operational constraints is to explicitly model unit commitment and operating reserves, as well as the lack of perfect foresight available to operators. While including these factors is not common in RA models for computational intensity reasons, recent work has identified that leaving out these considerations can bias RA metrics considerably (usually in the direction of overestimation) \cite{sun_insights_2022}. The ideal method for modeling these considerations is a rolling horizon stochastic optimization, which will be described further in the next section.

\subsection{Modeling energy storage}

The second important operational detail is the question of how to model operations of energy storage. As already identified, the most important consideration surrounding energy storage in RA assessments is to simulate chronological operations of the entire power system, in order to get an accurate picture of storage dispatch, as the availability of storage during challenging system hours will be dependent on its ability to charge and discharge outside of those hours. Ideally, the entire system would be simulated capturing all significant operational details, but for computational reasons some approximations, if carefully designed can be do an satisfactory job \cite{leibowicz_importance_2024}.

Even if chronological operations are simulated, however, there are still important modeling considerations surrounding energy storage in RA assessments. The key consideration is that energy storage operators in the real world do not have perfect foresight, limiting their ability to discharge when they are most needed. Therefore, if an RA assessment simulates system operations with perfect foresight, it is possible that the ability of storage to provide energy during challenging system hours could be overestimated.

There are multiple techniques that have been identified in the literature to handle this. One is to attempt to simulate the system using rolling horizon stochastic optimization, which reflects the foresight and decision environment that is actually available to grid participants \cite{stephen_impact_2022}. Another, less computationally demanding technique is to simulate system operations including storage with a relatively short, moving optimization ``window'' (e.g., only 48 hours at a time are simulated), paired with a penalty for not being at full charge at the end of each period. This limits parallelization as the ``windows'' must be simulated in series, but is very computationally efficient, and has been identified as a best practice for simulating realistic storage operations in RA assessments \cite{electric_power_research_institute_modeling_2023}.

\subsection{Modeling flexible load}

The third and final important operational detail we wish to touch on is the question of how to model flexible load. In most traditional RA assessments, load is treated as inelastic and static, and planning is based on building sufficient capacity to meet this static load. However, economics literature has identified that flexible load is a crucial piece of cost-effectively achieving high reliability \cite{borenstein_economics_2023, hogan_employing_2023, carden_economics_2011}. In addition, the emerging availability of smart meters and grid-connected devices such as smart appliances and electric vehicles means that load has the potential to become even more flexible and responsive. This raises the question of the best way to model this potential for flexibility in RA assessments.

Recent literature has identified several best practices for modeling flexible load in the context of RA assessments. In summary, a best practice for modeling flexible load consists of separating out static load from flexible load, and then characterizing flexible load in more detail using information such as elasticity and/or controllable load program details \cite{electric_power_research_institute_modeling_2023}. It is also very important to capture the weather-correlated nature of load in RA assessments. There are multiple ways of developing a load forecast in line with this theme, but for more details we refer the interested reader to recent reports that describe these methodologies in much more depth (primarily \cite{electric_power_research_institute_modeling_2023}). In the remainder of this section, rather, we focus on the modeling details that are important once a weather-correlated load forecast, with breakout into static vs flexible, has been developed.

Once the proportion of projected load that is flexible has been identified, it is necessary to identify the characteristics of the flexible load in order to simulate it. There are generally two types of flexible load: subscription programs in which consumers opt-in to have their load controlled within some boundaries, and voluntary price-based programs. These two types of flexible load have different modeling considerations. The former (subscription based) type can be modeled very similar to an energy-limited generator or storage device, with some operating parameters based on the characteristics of the program. The latter (price-based) type is more complicated as it depends on consumer behavior. Modeling price-based flexible loads requires an estimate of the elasticity of the demand, which can be difficult to estimate. An additional wrinkle is that demand elasticity can be highly dependent both on the level of demand, and on the time of year and system conditions. For example, if there is an extended heat wave, consumers might be responsive to price at the very beginning of it (more price elastic), but may be less willing to tolerate decreases in comfort as the heat wave goes on, leading to a lower price elasticity. Additionally, modeling price-based flexible loads of course also requires an economic dispatch simulation in order to come up with prices. For further details on modeling flexible loads, we refer the interested reader to \cite{electric_power_research_institute_modeling_2023}.

It is also important to note that, as far as modeling considerations go, we characterize customers on pre-specified TOU rates as static load and not flexible load. This is because customers on TOU rates are responding to pre-specified changes in TOU rates, not real time system conditions and prices, meaning that the responsiveness of these customers to real time prices does not need to be simulated. However, the rate design scheme of TOU customers should certainly be incorporated into the load forecast methodology used as an input to an RA assessment.

\subsection{Summary of operational details findings}

\begin{figure}[t]
  \centering
  \includegraphics[width=3.5in]{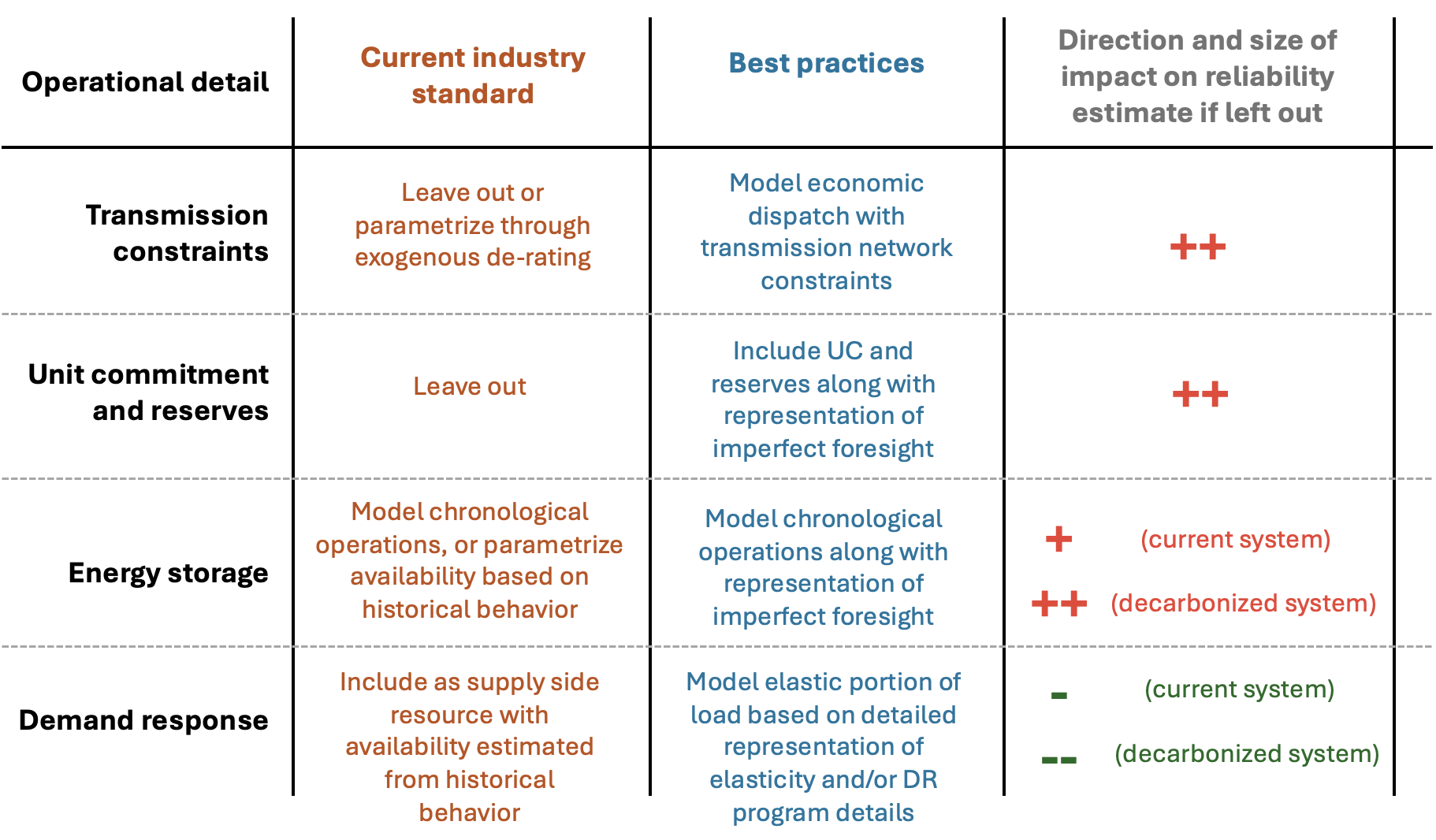}
  \caption{A schematic of the four categories of operational details discussed in this section, showing the current industry standard practices and best practices associated with each, along with the estimated direction and magnitude of impact on reliability if each is left out. A ``+'' in this last column means that leaving this detail out will lead to an overestimate of reliability.}
  \label{fig:operational_details}
\end{figure}

Figure \ref{fig:operational_details} summarizes the findings of this section, by showing the four categories of operational details discussed, along with current industry standard practices and best practices for each, as well as an estimate of the direction and magnitude of the impact on reliability if each is left out. This figure shows that reliability could be significantly overestimated in models that do not incorporate the best practices discussed, while this is balanced by inadequate modeling of demand response having the opposite direction of impact on reliability.

\section{Moving from assessment to procurement}
\label{procurement}

The sections above have been concerned with the \textbf{assessment} part of RA: i.e., the exercise of capturing the probabilistic characteristics of load shed events for a particular portfolio. An equally and perhaps more important component of RA is \textbf{procurement}: i.e. how to ensure that a particular power system has a desired level of reliability. Procurement is where actual investment decisions are made and this is driven by the RA assessment and by the particular procurement mechanism. Procurement practices differ for regions with different structures, but they share many commonalities, as the end goal is to procure a reliable portfolio of resources at least cost is the same. With this in mind we begin by describing reliability accreditation, which is a common need for any region, and then move to methods for regions with two different classes of procurement structures.

\subsection{Accreditation in procurement processes}

One key challenge in procuring a reliable portfolio is the determination of a ``capacity accreditation'' for resources, i.e. the need to capture each resource's contribution to RA so that a reliable overall portfolio can be developed. Such an accreditation methodology is generally required regardless of the portfolio procurement mechanism, although emerging methodologies may eliminate this need as described more below. Three examples of regulatory structures in which capacity accreditations are useful are:

\begin{itemize}
    \item \textbf{Capacity market.} When a capacity market is utilised, the accreditation is necessary to scale or weight the payments to generators in line with their contribution to RA \cite{mays_accreditation_2023}.
    \item \textbf{Strategic reserves.} In a system where a System Operator (SO) procures and operates out-of-market resources such as strategic reserves, the SO requires a transparent mechanism for comparing across different resources and compensating them appropriately \cite{neuhoff_coordinated_2016, finon_social_2008}.
    \item \textbf{Central planning.} In a system with centralized planning, ownership and operation of all generation assets, the accreditation may be useful, if not necessary, to inform generation and transmission planning models which cannot easily endogenize reliability considerations \cite{burdick_lighting_2022}.
\end{itemize}

In all of these contexts, there is a requirement for a robust estimate of how much capacity to ``credit'' a resource with, so that its contribution to RA can be accurately captured.

A well-established best practice for capacity accreditation in both of these contexts is the notion of Effective Load Carrying Capability or ELCC \cite{wang_crediting_2022,garver_effective_1966,keane_capacity_2011,zachary_integration_2022}. The ELCC for a resource, which can be expressed as a percentage of its nameplate capacity, is the equivalent amount of perfectly available capacity that results in the same improvement to system reliability as does adding the resource. Reliability in this context is generally quantified by either LOLP or EUE (both defined above). The rationale for using ELCC is that it is the metric that most accurately captures the reliability contribution of resources in a technology-neutral way. There is some debate in the literature on whether the marginal ELCC of a resource or average ELCC across multiple resource-portfolio combinations is the better metric to use for accreditation, but it is generally accepted that marginal ELCC is the best metric for accurately representing each resource's reliability contribution \cite{aagaard_marginal_2023}.

ELCC is calculated using the same models used for RA assessments (see above) for a particular portfolio; many existing models have this functionality built in. Therefore ELCC is as flawed as the RA assessment it is based on, but that is only part of its limitations.  A key limitation with ELCC is that for a given resource it is a function of both the buildout level of that particular resource, and also the buildout of all other resources in a portfolio. This functional dependence means that the ELCC of each resource is not a static value but rather one that depends on the final portfolio. This presents significant challenges that are described below.

\subsection{Operationalizing RA with central planning}

In regions where generation portfolios are centrally planned and procured, capacity expansion models (also known as generation and transmission expansion planing models) are generally used to make decisions about procurement. It historically been computationally difficult to endogenize reliability constraints directly in these models, necessitating an ``effective capacity'' constraint paired with a capacity accreditation for each resource, collectively referred to as a planning reserve margin constraint. As mentioned above, it is generally accepted that marginal ELCC is the ``right'' metric to use for accreditation in this context, although the dependence of ELCC on the penetration of all other resources must be captured. This problem can be solved through the construction of an ELCC surface that captures the key axes of functional dependence of resource ELCCs \cite{burdick_lighting_2022}.

It is important to note, however, that this methodology of using ELCC in a planning model is not the only possible methodology for endogenizing reliability in a planning process. One example of another possible methodology is using conditional value-at-risk metrics \cite{da_costa_reliability-constrained_2021}. Another possible methodology is to directly incorporate sampling of extreme conditions within a planning model, an area of active research \cite{stephen_enhanced_2022}. However, these methods are much less common in utility planning processes. For an example of a utility planning process using ELCC surfaces, see \cite{heath_elcc_2021}.

\subsection{Operationalizing RA without central planning}

In power systems where the generation portfolio is not centrally planned and procured but where there is some sort of capacity remuneration mechanism such as a capacity market or a backstop reliability reserve procurement mechanism, there is similarly a need for resource accreditation so that a cost-optimal reliable portfolio can be procured. This presents the question of how to best accredit resources' capacity contributions within such a framework. The question of capacity market design is extensively covered in the literature  and also includes considerations such as the formulation of capacity demand curves or other pricing curves that link to reliability (such as operating reserve demand curve in Texas), as well as the design of performance penalties \cite{hobbs_review_2001,billimoria_market_2019,mays_asymmetric_2019,duggan_capacity_2020}. Here, we focus only on the question of accreditation within a capacity mechanism.

In these non-centrally-planned contexts, marginal ELCC is still the ``right'' metric to use for capacity accreditation, but it is significantly more difficult to implement. Ideally, resources' capacity accreditations that are awarded could be a function of the final portfolio, but this conflicts with the need to have a stable and transparent accreditation for resources, as well as the need to have an accreditation ahead of a capacity auction so that resources can determine their optimal bids. This creates a chicken-or-the-egg problem that is to date unsolved. In practice, most markets that use marginal ELCC use a particular fixed portfolio as a baseline for marginal ELCC calculations, a simplification that allows firms to know their accreditation before bidding into capacity mechanisms \cite{noauthor_order_2022,noauthor_order_2024}.

It is important to note that there are also many other components of capacity market design that are necessary to encourage a competitive, transparent, and fair market. One of the most important considerations is the design of a performance penalty \cite{shu_beyond_2023}, so that resources are  incentivized to deliver on their accredited adequacy contribution. The list of capacity market design considerations is too long to thoroughly explain here; we instead refer the interested reader to the references at the beginning of this section.

\section{Synthesis and research directions}
\label{synthesis}

In this concluding section, we wish to highlight the key points made in this paper, and also present a framework for operationalizing the points made in the paper. Our key points are:

\begin{itemize}
    \item \textbf{Keeping the scope of RA manageable.} RA assessments can be a powerful tool for assessing and managing risk in the electric sector, and it is broad enough so as to encompass all of energy adequacy, flexibility adequacy, and capacity adequacy. However it is important to keep the scope of RA assessments limited to assessing the probabilistic risk of load shed events. Other types of reliability considerations such as operating reliability require entirely different modeling frameworks to assess and do not need to be included in the RA fold.
    \item \textbf{Understanding and managing different types of uncertainty.} Probabilistic RA assessments fundamentally can only handle types of uncertainty for which one can confidently assign probability distributions. Other types of uncertainty, known as deep uncertainty, are best managed through separate ``stress test'' analyses.
    \item \textbf{Including operational details.} There are many operational details, such as transmission constraints, energy storage operations, and flexible demand considerations, that are important to the accurate assessment of RA risk and should be included in RA assessments if feasible. However, including all of these considerations in an assessment is likely not feasible due to computational constraints.
    \item \textbf{Quantifying resources' reliability contributions in the context of procurement.} For procurement contexts where it is necessary to capture the reliability contribution of different resources, it is generally accepted that marginal ELCC is the ``right'' metric to use. However, emerging methodologies for procurement that do not require resource accreditations are an active area of research.
\end{itemize}

\begin{figure}[t]
  \centering
  \includegraphics[width=3.5in]{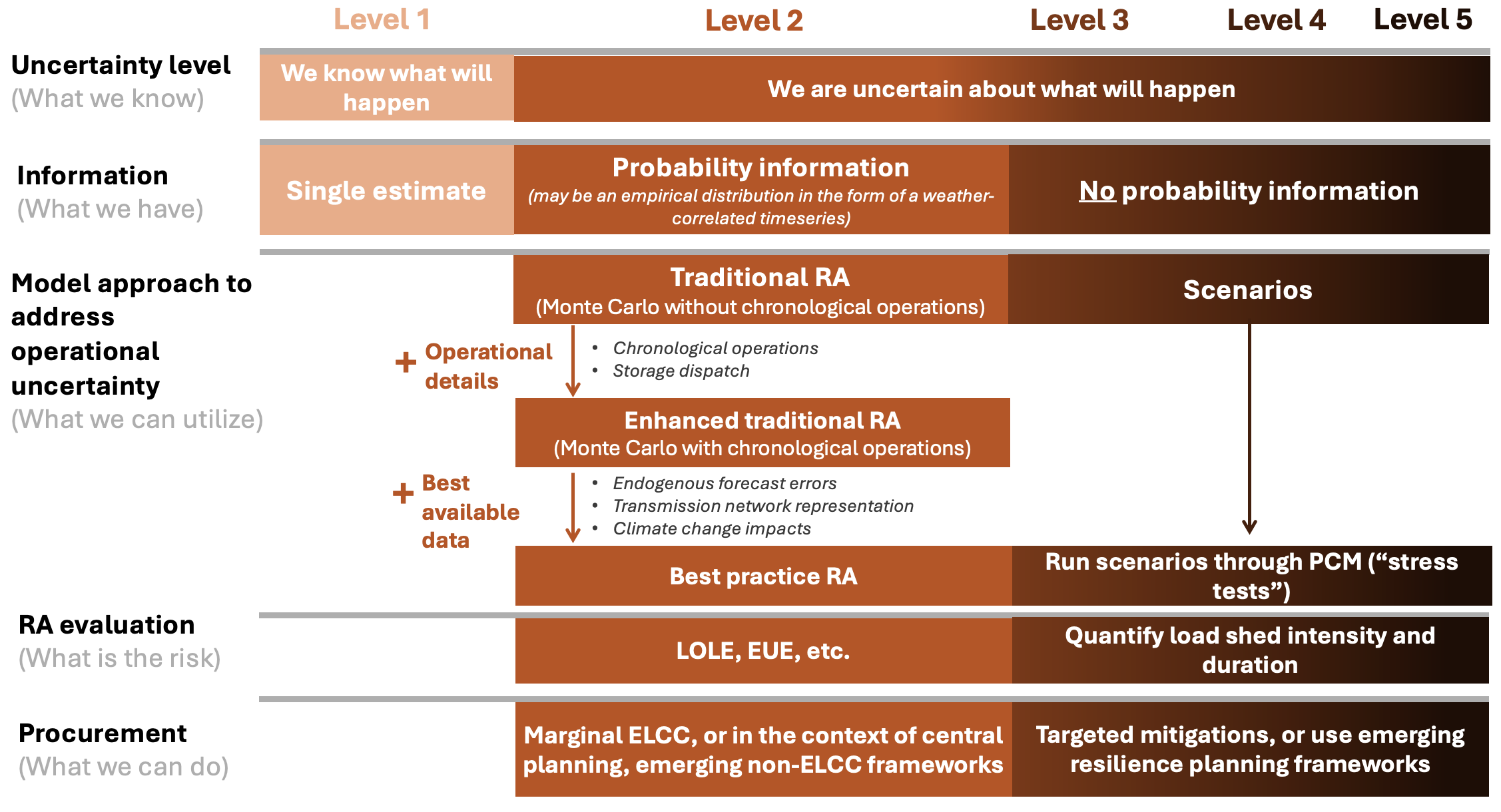}
  \caption{A schematic of the 5 levels of uncertainty, and how they relate to the RA assessment and procurement methodologies discussed in this paper.}
  \label{fig:framework_1}
\end{figure}

Figure \ref{fig:framework_1} provides a framework for thinking about the information presented in this paper, by showing the 5 levels of parametric uncertainty, and how they relate to the various modeling and procurement methodologies discussed. This figure shows that level 2 uncertainty, for which we have probabilities, is best dealt with through a Monte Carlo-based RA framework, ideally with sufficient representation of operational details, and is well suited to being paired with a procurement methodology that recognizes resource reliability contributions such as marginal ELCC. Level 3-5 uncertainty, on the other hand, for which we do not have probabilities but have some idea of what could happen, is best dealt with through scenario-based stress tests examining the impact of the events in question, and these stress tests can flow through to procurement based on a combination of targeted mitigations and/or emerging resilience planning frameworks, an area of active research.

\begin{figure}[t]
  \centering
  \includegraphics[width=3.5in]{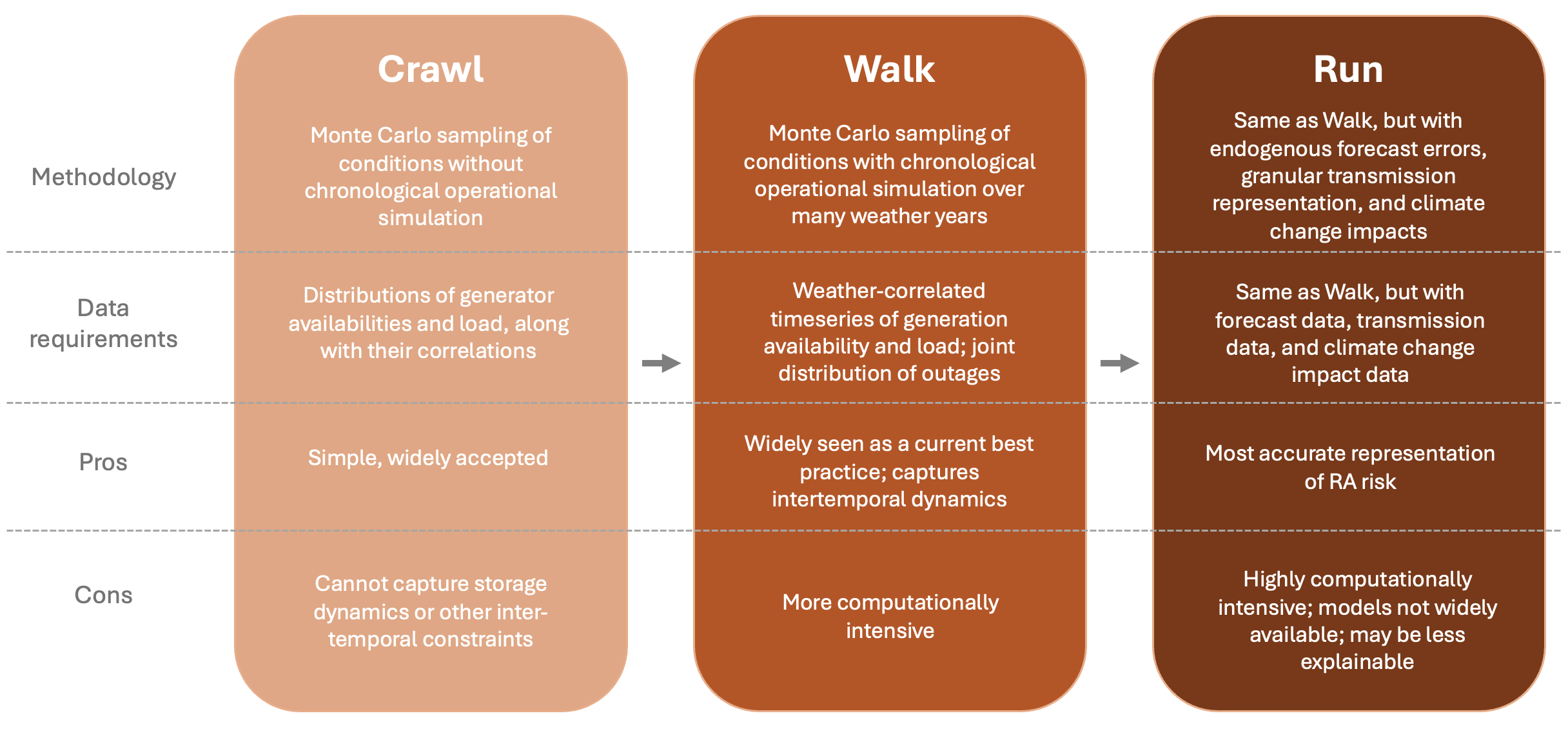}
  \caption{A depiction of three ``levels'' of advancement for RA assessment frameworks.}
  \label{fig:framework_2}
\end{figure}

Figure \ref{fig:framework_2} dives further into the question of how these concepts can be implemented for the assessment side specifically, by presenting a framework for thinking about levels of advancement for probabilistic RA assessments. We organize RA assessment practices into three ``levels'' of crawl, walk, and run, showing a progression from least detailed, but most common, to most detailed, but least common, illustrating how an entity needing to do reliability assessments could progress from where they are to where they could be, and highlighting the pros and cons of each stage. This figure shows that, while it is possible to describe what an ``ideal'' RA assessment framework looks like (as given by the Run column), the practices described in this category are not commonplace, and are highly computationally intensive.

Finally we wish to highlight several important directions for future research. First is the development of practical methodologies for including climate change considerations in reliability planning frameworks. Climate models provide some probabilistic information on future weather conditions, but these models themselves have uncertainty, which creates a challenge of how to appropriately model this uncertainty. Second is the further exploration of which operational details matter in RA assessments, particularly in the context of power systems with a high penetration of renewables and storage. Third is the development of methodologies for reliability-constrained planning that do not require a capacity accreditation based framework such as ELCC. Fourth is the development of capacity mechanisms that can endogenize the functional dependence of resource reliability contributions on the penetration of all other resources. With innovations in these areas, together with the adoption of cutting edge practices identified above, it is our belief that power systems around the world can stay reliable whilst supporting a transition to renewable energy.

\bibliographystyle{IEEEtran}
{\scriptsize \bibliography{bibliography.bib}}

\begin{thebibliography}{10}
\providecommand{\url}[1]{#1}
\csname url@samestyle\endcsname
\providecommand{\newblock}{\relax}
\providecommand{\bibinfo}[2]{#2}
\providecommand{\BIBentrySTDinterwordspacing}{\spaceskip=0pt\relax}
\providecommand{\BIBentryALTinterwordstretchfactor}{4}
\providecommand{\BIBentryALTinterwordspacing}{\spaceskip=\fontdimen2\font plus
\BIBentryALTinterwordstretchfactor\fontdimen3\font minus
  \fontdimen4\font\relax}
\providecommand{\BIBforeignlanguage}[2]{{%
\expandafter\ifx\csname l@#1\endcsname\relax
\typeout{** WARNING: IEEEtran.bst: No hyphenation pattern has been}%
\typeout{** loaded for the language `#1'. Using the pattern for}%
\typeout{** the default language instead.}%
\else
\language=\csname l@#1\endcsname
\fi
#2}}
\providecommand{\BIBdecl}{\relax}
\BIBdecl

\bibitem{ferc_february_2021}
FERC, NERC, and R.~Entities, ``The {February} 2021 {Cold} {Weather} {Outages}
  in {Texas} and the {South} {Central} {United} {States},'' FERC, NERC,
  Regional Entities, {FERC}, {NERC} and {Regional} {Entity} {Staff} {Report},
  Nov. 2021.

\bibitem{california_iso_root_2021}
C.~ISO, ``Root {Cause} {Analysis}: {Mid}-{August} 2020 {Extreme} {Heat}
  {Wave},'' California ISO, Tech. Rep., Jan. 2021.

\bibitem{cuff_europe_2023}
M.~Cuff, ``Europe survived its winter energy crisis, but what happens next
  year? {\textbar} {New} {Scientist},'' Mar. 2023.

\bibitem{shen_coping_2024}
B.~Shen, A.~Hove, J.~Hu, M.~Dupuy, L.~Bregnbæk, Y.~Zhang, and N.~Zhang,
  ``Coping with power crises under decarbonization: {The} case of {China},''
  \emph{Renewable and Sustainable Energy Reviews}, vol. 193, p. 114294, Apr.
  2024.

\bibitem{electric_power_research_institute_resource_2022-1}
E.~P.~R. Institute, ``Resource {Adequacy} {Philosophy}: {A} {Guide} to
  {Resource} {Adequacy} {Concepts} and {Approaches},'' Electric Power Research
  Institute, Technical {Update} 3002024368, Dec. 2022.

\bibitem{national_renewable_energy_laboratory_explained_2024}
N.~R.~E. Laboratory, ``\BIBforeignlanguage{en}{Explained: {Reliability} of the
  {Current} {Power} {Grid}},'' National Renewable Energy Laboratory, Tech.
  Rep., Jan. 2024.

\bibitem{north_american_electric_reliability_corporation_reliability_2013}
N.~A. E.~R. Corporation, ``Reliability {Terminology},'' North American Electric
  Reliability Corporation, Tech. Rep., Aug. 2013.

\bibitem{electric_power_research_institute_resource_2022}
{Electric Power Research Institute}, ``Resource {Adequacy} for a {Decarbonized}
  {Future}: {A} {Summary} of {Existing} and {Proposed} {Resource} {Adequacy}
  {Metrics},'' Electric Power Research Institute, Tech. Rep., 2022.

\bibitem{celebi_briefing_2024}
M.~Celebi, A.~Levitt, A.~W. Thompson, R.~Sreenath, X.~Bartone, S.~Willett, and
  H.~Ethier, ``\BIBforeignlanguage{en}{Briefing {Summary}: {Bulk} {System}
  {Reliability} for {Tomorrow}'s {Grid}},'' Brattle Group, Tech. Rep., Feb.
  2024.

\bibitem{lin_physically_2012}
N.~Lin, K.~Emanuel, M.~Oppenheimer, and E.~Vanmarcke,
  ``\BIBforeignlanguage{en}{Physically based assessment of hurricane surge
  threat under climate change},'' \emph{\BIBforeignlanguage{en}{Nature Climate
  Change}}, vol.~2, no.~6, pp. 462--467, Jun. 2012, publisher: Nature
  Publishing Group.

\bibitem{ayyub_climate-resilient_2018}
B.~M. Ayyub, ``Climate-{Resilient} {Infrastructure}: {Adaptive} {Design} and
  {Risk} {Management},'' ASCE, Tech. Rep., 2018.

\bibitem{hickford_review_2017}
A.~Hickford, S.~Blainey, A.~O. Hortelano, and R.~Pant, ``A {Review} of
  {Resilience} in {Interdependent} {Transport}, {Energy} and {Water}
  {Systems},'' The Resilience Shift, Tech. Rep., Feb. 2017.

\bibitem{stout_power_2019}
S.~Stout, N.~Lee, S.~Cox, J.~Elsworth, and J.~Leisch,
  ``\BIBforeignlanguage{en}{Power {Sector} {Resilience} {Planning} {Guidebook}:
  {A} {Self}-{Guided} {Reference} for {Practitioners}},'' National Renewable
  Energy Laboratory, Tech. Rep. NREL/TP--7A40-73489, 1529875, Jun. 2019.

\bibitem{fang_adaptive_2019}
Y.-P. Fang and E.~Zio, ``An adaptive robust framework for the optimization of
  the resilience of interdependent infrastructures under natural hazards,''
  \emph{European Journal of Operational Research}, vol. 276, no.~3, pp.
  1119--1136, Aug. 2019.

\bibitem{belle_resilience-based_2023}
A.~Bellè, A.~F. Abdin, Y.-P. Fang, Z.~Zeng, and A.~Barros, ``A
  resilience-based framework for the optimal coupling of interdependent
  critical infrastructures,'' \emph{Reliability Engineering \& System Safety},
  vol. 237, p. 109364, Sep. 2023.

\bibitem{national_academies_of_sciences_engineering_and_medicine_enhancing_2017}
{National Academies of Sciences, Engineering, and Medicine}, \emph{Enhancing
  the {Resilience} of the {Nation}'s {Electricity} {System}}.\hskip 1em plus
  0.5em minus 0.4em\relax Washington, D.C.: National Academies Press, Sep.
  2017.

\bibitem{kasina_resilience_2024}
S.~Kasina, N.~Schlag, A.~Olson, and R.~Orans,
  ``\BIBforeignlanguage{en}{Resilience in planning: {Its} relationship to
  {Reliability} and a practical implementation guide},'' Energy and
  Environmental Economics, Tech. Rep., 2024.

\bibitem{cainey_resilience_2019}
J.~M. Cainey, ``\BIBforeignlanguage{en}{Resilience and reliability for
  electricity networks},'' \emph{\BIBforeignlanguage{en}{Proceedings of the
  Royal Society of Victoria}}, vol. 131, no.~1, pp. 44--52, 2019, publisher:
  CSIRO PUBLISHING.

\bibitem{cox_jr_confronting_2012}
L.~A.~T. Cox~Jr., ``\BIBforeignlanguage{en}{Confronting {Deep} {Uncertainties}
  in {Risk} {Analysis}},'' \emph{\BIBforeignlanguage{en}{Risk Analysis}},
  vol.~32, no.~10, pp. 1607--1629, 2012, \_eprint:
  https://onlinelibrary.wiley.com/doi/pdf/10.1111/j.1539-6924.2012.01792.x.

\bibitem{thissen_public_2013}
W.~A.~H. Thissen and W.~E. Walker, Eds., \emph{\BIBforeignlanguage{en}{Public
  {Policy} {Analysis}: {New} {Developments}}}, ser. International {Series} in
  {Operations} {Research} \& {Management} {Science}.\hskip 1em plus 0.5em minus
  0.4em\relax Boston, MA: Springer US, 2013, vol. 179.

\bibitem{stenclik_redefining_2021}
D.~Stenclik, A.~Bloom, W.~Cole, A.~Acevedo, G.~Stephen, and A.~Tuohy,
  ``\BIBforeignlanguage{en}{Redefining {Resource} {Adequacy} for {Modern}
  {Power} {Systems}: {A} {Report} of the {Redefining} {Resource} {Adequacy}
  {Task} {Force}},'' Energy Systems Integration Group, Tech. Rep.
  NREL/TP--5C00-80896, 1961567, MainId:78674, Jan. 2021.

\bibitem{lin_equity-based_2022}
Y.~Lin, J.~Wang, and M.~Yue, ``Equity-based grid resilience: {How} do we get
  there?'' \emph{The Electricity Journal}, vol.~35, no.~5, p. 107135, Jun.
  2022.

\bibitem{pfeifenberger_resource_2013}
J.~P. Pfeifenberger, K.~Spees, K.~Carden, and N.~Wintermantel,
  ``\BIBforeignlanguage{en}{Resource {Adequacy} {Requirements}: {Reliability}
  and {Economic} {Implications}},'' Federal Energy Regulatory Commission, Tech.
  Rep., Sep. 2013.

\bibitem{leibowicz_importance_2024}
B.~D. Leibowicz, N.~Zhang, J.~P. Carvallo, P.~H. Larsen, T.~Carr, and S.~Baik,
  ``The importance of capturing power system operational details in resource
  adequacy assessments,'' \emph{Electric Power Systems Research}, vol. 228, p.
  110057, Mar. 2024.

\bibitem{electric_power_research_institute_modeling_2023}
{Electric Power Research Institute}, ``Modeling {New} and {Existing}
  {Technologies} and {System} {Components} in {Resource} {Adequacy},'' Electric
  Power Research Institute, Tech. Rep. 3002027830, 2023.

\bibitem{sun_insights_2022}
Y.~Sun, B.~Frew, S.~Dalvi, and S.~C. Dhulipala, ``Insights into methodologies
  and operational details of resource adequacy assessment: {A} case study with
  application to a broader flexibility framework,'' \emph{Applied Energy}, vol.
  328, p. 120191, Dec. 2022.

\bibitem{nolan_capacity_2017}
S.~Nolan, O.~Neu, and M.~O’Malley, ``Capacity value estimation of a
  load-shifting resource using a coupled building and power system model,''
  \emph{Applied Energy}, vol. 192, pp. 71--82, Apr. 2017.

\bibitem{stephen_impact_2022}
G.~Stephen, T.~Joswig-Jones, S.~Awara, and D.~Kirschen, ``Impact of {Storage}
  {Dispatch} {Assumptions} on {Resource} {Adequacy} and {Capacity} {Credit},''
  in \emph{2022 17th {International} {Conference} on {Probabilistic} {Methods}
  {Applied} to {Power} {Systems} ({PMAPS})}, Jun. 2022, pp. 1--6, iSSN:
  2642-6757.

\bibitem{borenstein_economics_2023}
S.~Borenstein, J.~Bushnell, and E.~Mansur, ``\BIBforeignlanguage{en}{The
  {Economics} of {Electricity} {Reliability}},''
  \emph{\BIBforeignlanguage{en}{Journal of Economic Perspectives}}, vol.~37,
  no.~4, pp. 181--206, Dec. 2023.

\bibitem{hogan_employing_2023}
M.~Hogan, ``\BIBforeignlanguage{en}{Employing {Price}-{Responsive} {Demand} to
  {Reduce} the {Investment} {Challenge}},'' ESIG, Tech. Rep., Jan. 2023.

\bibitem{carden_economics_2011}
K.~Carden, N.~Wintermantel, and J.~P. Pfeifenberger,
  ``\BIBforeignlanguage{en}{The {Economics} of {Resource} {Adequacy}
  {Planning}: {Why} {Reserve} {Margins} {Are} {Not} {Just} {About} {Keeping}
  the {Lights} {On}},'' National Regulatory Research Institute, Tech. Rep.,
  2011.

\bibitem{mays_accreditation_2023}
J.~Mays, ``\BIBforeignlanguage{en-US}{Accreditation, {Performance}, and
  {Credit} {Risk} in {Electricity} {Capacity} {Markets}},'' Jun. 2023.

\bibitem{neuhoff_coordinated_2016}
K.~Neuhoff, J.~Diekmann, F.~Kunz, S.~Rüster, W.-P. Schill, and S.~Schwenen,
  ``A coordinated strategic reserve to safeguard the {European} energy
  transition,'' \emph{Utilities Policy}, vol.~41, pp. 252--263, Aug. 2016.

\bibitem{finon_social_2008}
D.~Finon, G.~Meunier, and V.~Pignon, ``The social efficiency of long-term
  capacity reserve mechanisms,'' \emph{Utilities Policy}, vol.~16, no.~3, pp.
  202--214, Sep. 2008.

\bibitem{burdick_lighting_2022}
A.~Burdick, N.~Schlag, A.~Au, R.~Go, Z.~Ming, and A.~Olson, ``Lighting a
  {Reliable} {Path} to 100\% {Clean} {Electricity}: {Evolving} {Resource}
  {Adequacy} {Practices} for a {Decarbonizing} {Grid},'' \emph{IEEE Power and
  Energy Magazine}, vol.~20, no.~4, pp. 30--43, Jul. 2022, conference Name:
  IEEE Power and Energy Magazine.

\bibitem{wang_crediting_2022}
S.~Wang, N.~Zheng, C.~D. Bothwell, Q.~Xu, S.~Kasina, and B.~F. Hobbs,
  ``Crediting {Variable} {Renewable} {Energy} and {Energy} {Storage} in
  {Capacity} {Markets}: {Effects} of {Unit} {Commitment} and {Storage}
  {Operation},'' \emph{IEEE Transactions on Power Systems}, vol.~37, no.~1, pp.
  617--628, Jan. 2022, conference Name: IEEE Transactions on Power Systems.

\bibitem{garver_effective_1966}
L.~L. Garver, ``Effective {Load} {Carrying} {Capability} of {Generating}
  {Units},'' \emph{IEEE Transactions on Power Apparatus and Systems}, vol.
  PAS-85, no.~8, pp. 910--919, Aug. 1966, conference Name: IEEE Transactions on
  Power Apparatus and Systems.

\bibitem{keane_capacity_2011}
A.~Keane, M.~Milligan, C.~J. Dent, B.~Hasche, C.~D'Annunzio, K.~Dragoon,
  H.~Holttinen, N.~Samaan, L.~Soder, and M.~O'Malley, ``Capacity {Value} of
  {Wind} {Power},'' \emph{IEEE Transactions on Power Systems}, vol.~26, no.~2,
  pp. 564--572, May 2011, conference Name: IEEE Transactions on Power Systems.

\bibitem{zachary_integration_2022}
S.~Zachary, A.~Wilson, and C.~Dent, ``\BIBforeignlanguage{en}{The {Integration}
  of {Variable} {Generation} and {Storage} into {Electricity} {Capacity}
  {Markets}},'' \emph{\BIBforeignlanguage{en}{The Energy Journal}}, vol.~43,
  no.~4, pp. 231--250, May 2022, publisher: SAGE Publications.

\bibitem{aagaard_marginal_2023}
T.~Aagaard and A.~N. Kleit, ``Marginal vs. average effective load carrying
  capability: {How} should capacity markets deal with alternative generation
  forms?'' \emph{Utilities Policy}, vol.~84, p. 101654, 2023, publisher:
  Elsevier.

\bibitem{da_costa_reliability-constrained_2021}
L.~C. da~Costa, F.~S. Thomé, J.~D. Garcia, and M.~V.~F. Pereira,
  ``Reliability-{Constrained} {Power} {System} {Expansion} {Planning}: {A}
  {Stochastic} {Risk}-{Averse} {Optimization} {Approach},'' \emph{IEEE
  Transactions on Power Systems}, vol.~36, no.~1, pp. 97--106, Jan. 2021,
  conference Name: IEEE Transactions on Power Systems.

\bibitem{stephen_enhanced_2022}
G.~Stephen and D.~Kirschen, ``Enhanced {Representations} of {Thermal}
  {Generator} {Outage} {Risk} in {Capacity} {Expansion} {Models},'' in
  \emph{2022 17th {International} {Conference} on {Probabilistic} {Methods}
  {Applied} to {Power} {Systems} ({PMAPS})}, Jun. 2022, pp. 1--6, iSSN:
  2642-6757.

\bibitem{heath_elcc_2021}
B.~Heath and J.~Nelson, ``{ELCC} {Surface} in {Resource} {Planning}: {NV}
  {Energy} {IRP},'' 2021.

\bibitem{hobbs_review_2001}
B.~F. Hobbs, J.~Iñón, and M.~Kahal, ``A {Review} of {Issues} {Concerning}
  {Electric} {Power} {Capacity} {Markets},'' Maryland Department of Natural
  Resources, Project {Report}, 2001.

\bibitem{billimoria_market_2019}
F.~Billimoria and R.~Poudineh, ``Market design for resource adequacy: {A}
  reliability insurance overlay on energy-only electricity markets,''
  \emph{Utilities Policy}, vol.~60, p. 100935, Oct. 2019.

\bibitem{mays_asymmetric_2019}
J.~Mays, D.~P. Morton, and R.~P. O’Neill,
  ``\BIBforeignlanguage{en}{Asymmetric risk and fuel neutrality in electricity
  capacity markets},'' \emph{\BIBforeignlanguage{en}{Nature Energy}}, vol.~4,
  no.~11, pp. 948--956, Nov. 2019, publisher: Nature Publishing Group.

\bibitem{duggan_capacity_2020}
J.~E. Duggan, ``\BIBforeignlanguage{en}{Capacity {Market} {Mechanism}
  {Analyses}: a {Literature} {Review}},'' \emph{\BIBforeignlanguage{en}{Current
  Sustainable/Renewable Energy Reports}}, vol.~7, no.~4, pp. 186--192, Dec.
  2020.

\bibitem{noauthor_order_2022}
``Order {Accepting} {Tariff} {Revisions} {Subject} {To} {Condition} [{FERC}
  {Decision} regarding {NYISO} {Capacity} {Accreditation}],'' May 2022.

\bibitem{noauthor_order_2024}
``Order {Accepting} {Tariff} {Revisions} {Subject} {To} {Condition} [{FERC}
  {Decision} regarding {PJM} {Capacity} {Accreditation}],'' Jan. 2024.

\bibitem{shu_beyond_2023}
H.~Shu and J.~Mays, ``Beyond capacity: {Contractual} form in electricity
  reliability obligations,'' \emph{Energy Economics}, vol. 126, p. 106943, Oct.
  2023.

\end{thebibliography}

\vfill

\end{document}